\documentclass[cameraready]{Interspeech}
\usepackage[utf8]{inputenc}
\usepackage{booktabs}
\usepackage{amssymb}
\usepackage{amsmath}
\usepackage{microtype}
\usepackage{newunicodechar}
\newunicodechar{，}{,}
\title{Visually-Guided Spatial Audio Generation for 360$^\circ$ In-the-Wild Speech Scenes}

\author{Qingyu}{Luo}
\author{Peng}{Zhang}
\author{Wenwu}{Wang}
\author{Philip J.B.}{Jackson}

\address{
    Centre for Vision, Speech and Signal Processing (CVSSP), University of Surrey, U.K.
}

\email{\{qingyu.luo, p.zhang, w.wang, p.jackson\}@surrey.ac.uk}

\keywords{Spatial Audio, Ambisonics, $360^\circ$ Video, Speech Spatialization}

\usepackage{comment}
\usepackage{makecell}
\newunicodechar{́}{\'}

\begin{document}

\maketitle

\begin{abstract}
   Spatial audio is a key component of immersive $360^\circ$ media, yet high-quality spatial capture remains limited in real-world speech-dominant scenes. We study visually guided First-Order Ambisonics (FOA) speech spatialization in the wild: given aligned $360^\circ$ video and an omnidirectional audio track, we recover the missing directional FOA components.
   To support this task, we introduce YT-SPEECH, a speech-oriented $360^\circ$ video-FOA dataset curated from YouTube. We propose a two-stage Localizer-Renderer framework, where an audio-visual segmentation backbone provides frame-wise spatial heatmaps and a conditional complex-domain U-Net reconstructs directional FOA signals from the omnidirectional channel. A confidence-based gating strategy stabilizes conditioning under ambiguous acoustic conditions. Experiments show improved reconstruction fidelity, spatial accuracy, and perceptual speech quality relative to ablated variants and prior approaches.
\end{abstract}


\section{Introduction}

Immersive media such as $360^\circ$ video and virtual reality are increasingly deployed in telepresence, panoramic platforms, and interactive content.
To sustain immersion, spatial audio should remain spatially coherent with the captured visual scene during playback.
Speech-centric scenes are particularly common in real-world $360^\circ$ videos (e.g., interviews, conversations, commentary), and localization errors on foreground speech are particularly noticeable  compared to diffuse background sounds.
These factors motivate visually guided speech spatialization in the wild, where the goal is to recover a coherent spatial impression of speech from unconstrained recordings.

Most existing speech spatial-audio research renders mono speech into binaural or stereo signals~\cite{Binaural,NFS,TTMBA}, sometimes with additional conditioning such as geometry or text~\cite{Asaudio}.
However, fixed binaural outputs are tied to a listener orientation, whereas First-Order Ambisonics (FOA) is a scene-based representation that can be rotated and decoded into multiple formats, including head-tracked binaural.
As summarized in Table~\ref{tab:dataset}, available speech spatial datasets are predominantly binaural, limited in scale, and often rely on controlled geometric labels; FOA speech datasets, when available, are typically simulated and rarely paired with panoramic video cues~\cite{LibriSpeech}.
This mismatch in format, scale, and visual grounding significantly limits the study of visually guided FOA speech spatialization in real-world scenes.
\begin{figure}[t]
  \centering
  \includegraphics[width=1\linewidth]{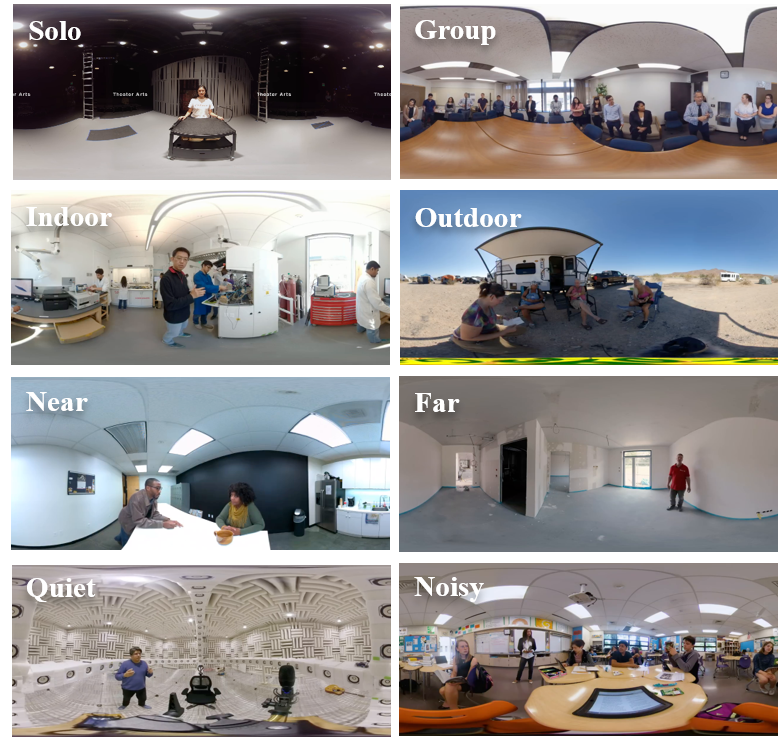}
  \caption{Sample scenes from the proposed YT-SPEECH.}
  \label{fig:samples}
\end{figure}

In parallel, $360^\circ$ video-to-FOA generation has progressed rapidly.
Existing approaches broadly fall into two paradigms: explicit spatial reconstruction and end-to-end generation.
Among explicit reconstruction methods, SpatialAudioGen (SAG)~\cite{360Video2018} and its extension~\cite{panningloss2024} remain among the few works addressing video-conditioned FOA reconstruction from an omnidirectional input.
They adopt a self-supervised decomposition pipeline that separates latent tracks and estimates per-track spatial parameters without source-level supervision.
However, such label-free separation does not guarantee clean source disentanglement and may introduce artefacts, thereby degrading the reconstructed directional channels.
Rana et al.~\cite{rana2019towards} propose a zero-shot direction-first pipeline based on analytic ambisonic encoding, which is interpretable but may be less expressive in reverberant or diffuse environments.
In contrast, large-scale end-to-end models such as OmniAudio~\cite{Omniaudio2025} and ViSAGe~\cite{ViSAGe2025} learn open-domain video-to-spatial-audio generation directly from data, emphasizing semantic consistency rather than accurate spatial reconstruction.

\begin{figure*}[!t]
  \centering
  \includegraphics[width=0.95\textwidth]{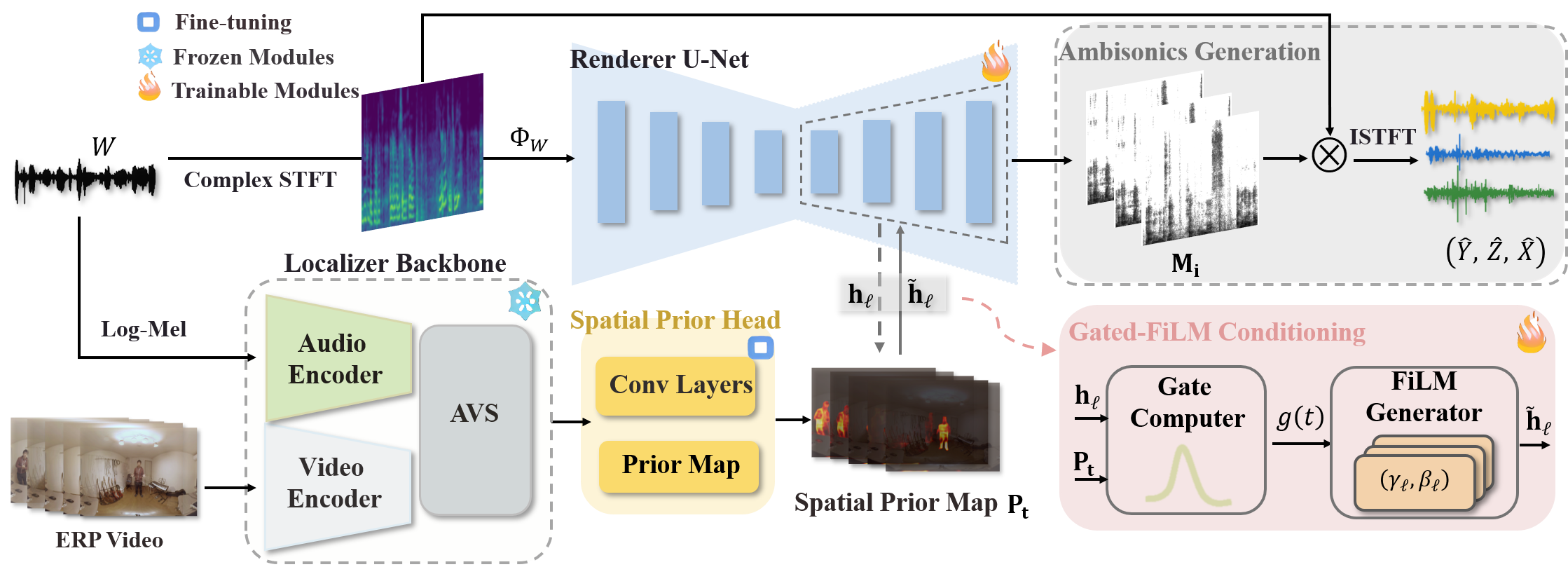}
  \caption{Overview of the proposed Localizer–Renderer framework for visually guided FOA spatial reconstruction.}
  \label{fig:network}
  \vspace{-10pt}
\end{figure*}

Following the direction-first strategy, we focus on explicit spatial reconstruction in speech-dominant $360^\circ$ scenes.
Reliable localization in unconstrained panoramic videos remains challenging under limited supervision: sound event localization models often rely on coarse spatial grids or discrete direction-of-arrival (DOA) estimation, limiting their suitability for visually grounded directional reconstruction in panoramic scenes, whereas speaker-centric pipelines~\cite{speakerdetection} depend on cascaded face detection and active-speaker association, which may not generalize well to in-the-wild $360^\circ$ content.
To obtain dense and visually grounded spatial cues without explicit geometric supervision, we adopt an Audio-Visual Segmentation (AVS) framework to produce audio-conditioned spatial activations that serve as an interpretable prior for directional reconstruction.

In this paper, we study explicit spatial reconstruction for speech-dominant $360^\circ$ video scenes, where directional FOA components are recovered from aligned video and an omnidirectional audio input.
To support this setting, we curate YT-SPEECH, a high-quality speech-oriented dataset built from publicly available $360^\circ$ videos with spatial audio.
We further propose a Localizer--Renderer framework that leverages dense audio-visual spatial priors and performs phase-consistent reconstruction in the complex spectral domain. Demo samples are available at: \url{https://spatial-audio-demo.github.io/demos/}.

The key contributions of our work are summarized as follows: 
\begin{itemize}
        \item YT-SPEECH, to our knowledge, the first speech-oriented $360^\circ$ video–FOA dataset designed for visually guided spatial reconstruction.
        \item  A Localizer--Renderer framework for direction-consistent FOA reconstruction using audio-visual priors.
        \item  Comprehensive quantitative and qualitative evaluations demonstrating robustness in reconstruction fidelity, spatial accuracy, and perceptual speech quality.
\end{itemize}

\section{Dataset}
Existing speech spatial-audio datasets mostly provide binaural recordings in controlled settings or simulated FOA speech with text/geometry supervision, while paired $360^\circ$ video–FOA speech data in the wild remains scarce. This limitation is particularly restrictive for video-guided FOA spatialization, where reliable on-screen speech cues and audio quality are crucial. To address this gap, we curate YT-SPEECH, a speech-oriented dataset built from publicly available YouTube $360^\circ$ videos with FOA audio. 

To ensure high-quality speech and visually grounded samples, we implement a multi-stage filtering pipeline. 
For spatial audio quality, we discard clips with invalid multi-channel layouts or abnormal channel-energy patterns~\cite{energy_critera}, retaining only correctly formatted FOA recordings with consistent per-channel activity. 
For speech selection, we detect speech-active regions using speaker diarization\footnote{\url{https://github.com/pyannote/pyannote-audio}} and verify the segments with Whisper ASR~\cite{whisper}; an AudioSet-based classifier~\cite{panns} excludes clips dominated by music or singing so that retained clips remain speech-dominant. 
To ensure consistent visual presence of speakers, we detect on-screen persons using the YOLOv8-nano model\footnote{\url{https://github.com/ultralytics/ultralytics}} and retain segments with visible speakers, requiring both a high frame-wise presence ratio and a minimum bounding-box area ratio to favour visible talkers over distant or voice-over-only cases. 
We further score audio--visual correspondence using a learned alignment criterion and remove clips with low consistency scores. Finally, we manually inspect the remaining samples and remove background narration or voice-over cases where speech is not visually grounded.

\begin{table}[t]
  \centering
  \caption{Comparison of speech spatial audio datasets. Bin/FOA denote binaural and first-order ambisonics. Sim/Rec/Crawled denote simulated, recorded and web-crawled data.}
  \label{tab:dataset}
  \footnotesize
  \setlength{\tabcolsep}{3pt}
  \renewcommand{\arraystretch}{1.2} 
  \resizebox{\columnwidth}{!}{%
  \begin{tabular}{lccccc}
    \hline
    \textbf{Dataset} & \textbf{Hours} &  \textbf{Format} & \textbf{Source} & \textbf{Label}   \\
    \hline
   Spatial LibriSpeech~\cite{LibriSpeech} & 650 & FOA & Sim & Text/Geometric \\
    EasyCom~\cite{easycom} & 5 & Bin & Rec & Text/Geometric \\
    Binaural~\cite{Binaural} & 2 & Bin & Rec & Geometric\\
    MRSDrama~\cite{MRSDrama} & 98 & Bin & Rec & Text/Geometric/Video \\
    \hline
    \textbf{YT-SPEECH} & 8.9 & FOA  & Crawled & $360^\circ$ Video  \\
    \hline
  \end{tabular}}
\end{table}

The final dataset contains 8.9~hours of 5-second clips at 24~kHz from 197 source videos, split by video IDs into training, validation, and test sets to prevent content leakage. The dataset spans diverse real-world speech scenes, including indoor and outdoor environments, solo and group interactions, near- and far-field speech, and varying background noise conditions. Example clips are shown in Fig.~\ref{fig:samples}.


\section{Method}
Given an aligned $360^\circ$ video sequence $\mathbf{V}$ and the omnidirectional FOA channel $W$, we reconstruct the missing directional components $(Y,Z,X)$ in the complex short-time Fourier transform (STFT) domain to obtain a coherent FOA representation. Let $\Phi_i$ denote the complex spectrum obtained via the STFT of channel $i\in\{W,Y,Z,X\}$. Our model preserves $\Phi_W$ and predicts $(\hat{\Phi}_Y,\hat{\Phi}_Z,\hat{\Phi}_X)$, forming $\hat{\Phi}_{\mathrm{FOA}}=[\Phi_W;\allowbreak \hat{\Phi}_Y;\allowbreak \hat{\Phi}_Z;\allowbreak \hat{\Phi}_X]$. An overview of the framework is shown in Fig.~\ref{fig:network}.

\subsection{Localizer}
\label{sec:tables}

To guide directional synthesis, the Localizer produces a dense, audio-conditioned spatial heatmap over the full $360^\circ$ field of view. Since panoramic video is represented in equirectangular projection (ERP), preserving the $2{:}1$ geometry is critical for maintaining spatial correspondence.

We adopt AVS~\cite{tpavi} as the backbone, which consists of a dedicated audio encoder~\cite{VGGlike} and a video encoder~\cite{pvt} with cross-modal interaction. The ERP frame $\mathbf{V}\in\mathbb{R}^{3\times224\times448}$ and synchronized $W$ audio are processed by the frozen backbone to extract a shared audio–visual feature map. Instead of segmentation masks, we repurpose the original mask head as a Spatial Prior Head and fine-tune it to produce dense spatial activation $\mathbf{S}_p$. Circular padding is applied to preserve horizontal wrap-around continuity in ERP space.
The activation is transformed into a normalized spatial prior:
\begin{equation}
\mathbf{P}_t
=
\mathrm{Norm}\big(
\operatorname{Pool}(\mathrm{softplus}(\mathbf{S}_p))
\big),
\end{equation}
where $\mathrm{softplus}(\cdot)$ ensures non-negativity, $\operatorname{Pool}(\cdot)$ aggregates activations across the spatial grid, and $\mathrm{Norm}(\cdot)$ normalizes the map so that all entries sum to one. The resulting prior satisfies $\mathbf{P}_t\in\mathbb{R}^{7\times14}$ with $\sum_i \mathbf{P}_{t,i}=1$, forming a spatial distribution over the panoramic field of view.

The frame-level priors are temporally aligned to the STFT resolution via interpolation and renormalized to maintain probabilistic consistency.

\subsection{Renderer}
The Renderer synthesizes the directional spectra $(\hat{\Phi}_Y,\hat{\Phi}_Z,\hat{\Phi}_X)$ from the omnidirectional spectrum $\Phi_W$ conditioned on the spatial prior $\mathbf{P}_t$. We adopt a U-Net-style backbone~\cite{ambi_unet} operating in the complex STFT domain, enabling phase-sensitive spectral transformation while preserving the cross-channel coherence required for FOA rendering. Consistent with the Spatial Prior Head, circular padding~\cite{circlepadding} is employed in convolutional layers to respect ERP wrap-around continuity.

The prior $\mathbf{P}_t$ provides frame-wise spatial guidance but may be ambiguous in diffuse or visually uncertain moments. 
We therefore compute a scalar confidence gate $g(t)\in[0,1]$ from the flattened prior $\mathbf{p}_t\in\mathbb{R}^K$, where $t$ indexes the STFT time frame and $K$ is the number of spatial bins. 
Confidence is decomposed into two complementary factors: a peak-based measure capturing concentration and an entropy-based measure reflecting uncertainty:
\begin{equation}
g_{\max}(t)
=
\frac{\max_i p_{t,i} - 1/K}{1 - 1/K},
\end{equation}
\begin{equation}
g_H(t)
=
1 - \frac{H(\mathbf{p}_t)}{\log K},
\end{equation}
where $H(\mathbf{p}_t) = -\sum_{i=1}^{K} p_{t,i}\log p_{t,i}$. 
The final confidence is defined as $g(t) = g_{\max}(t)\, g_H(t)$, so that conditioning strength increases when the spatial prior is concentrated and has low-entropy.

A lightweight prior adapter maps $\mathbf{P}_t$ to modulation parameters $(\boldsymbol{\gamma}_\ell, \boldsymbol{\beta}_\ell)$ for each decoder block $\ell$. Let $\mathbf{h}_\ell$ denote the intermediate feature map at decoder block $\ell$ of the Renderer U-Net. 
We apply gated feature-wise linear modulation (FiLM):
\begin{equation}
\tilde{\mathbf{h}}_\ell
=
\mathbf{h}_\ell
\odot
\bigl(1 + g(t)\boldsymbol{\gamma}_\ell\bigr)
+
g(t)\boldsymbol{\beta}_\ell,
\end{equation}
where $\odot$ denotes element-wise multiplication. 
The modulated feature $\tilde{\mathbf{h}}_\ell$ is used to condition subsequent decoder features.  This mechanism adaptively balances visual guidance and audio evidence according to the reliability of the spatial prior.

Instead of using explicit DOA-based encoding~\cite{rana2019towards}, 
the Renderer learns a direction-dependent complex spectral projection.  Specifically, it predicts complex-valued masks $\mathbf{M}_i \in \mathbb{R}^{2\times F\times T}$ 
that are applied element-wise to $\Phi_W$ to obtain each directional spectrum, i.e., $\hat{\Phi}_{i}=\mathbf{M}_{i}\odot\Phi_{W}$ for $i\in\{Y,Z,X\}$. 
This formulation preserves the spectral structure of $W$ while enabling direction-specific shaping required for FOA reconstruction.

We optimize a confidence-weighted combination of spectral and waveform losses:
\begin{equation}
\mathcal{L}
=
\frac{1}{|\Omega|}
\sum_{b,c,f,t}
w(t)
\Big(
\lambda_1 \mathcal{L}_{\mathrm{MRS}}
+
\lambda_2 \mathcal{L}_{\mathrm{mag}}
+
\lambda_3 \mathcal{L}_{\ell_2}
\Big),
\label{eq:loss}
\end{equation}
where $\Omega$ indexes batch, channel, frequency, and time dimensions. 
The frame-wise confidence weight is defined as $w(t)=\alpha+(1-\alpha)g(t)$ with $\alpha\in[0,1]$, where $g(t)$ denotes the prior confidence. 
$\mathcal{L}_{\mathrm{MRS}}$ is the multi-resolution STFT loss~\cite{MRSTFT}, 
$\mathcal{L}_{\mathrm{mag}}$ enforces magnitude consistency in the STFT domain, 
and $\mathcal{L}_{\ell_2}$ is the waveform-level $\ell_2$ loss.


\section{Experiments}

\begin{table*}[t]
  \centering
  \caption{\textit{Localizer ablation and Renderer comparison under the Localizer–Renderer framework on the YT-SPEECH dataset. The best score for each metric is marked with an asterisk ($\cdot^{\ast}$) and the second best score is marked with a dagger ($\cdot^{\dagger}$).}}
  \label{tab:foa}
  \footnotesize
  \setlength{\tabcolsep}{5pt}
  \begin{tabular*}{\textwidth}{@{\extracolsep{\fill}}lllll lll lll}
    \toprule
    \multicolumn{1}{c}{\textbf{Model}} &
    \multicolumn{4}{c}{\textbf{Reconstruction Metrics}} &
    \multicolumn{3}{c}{\textbf{Spatial Metrics}} &
    \multicolumn{3}{c}{\textbf{Speech Metrics}}\\
    \cmidrule(lr){2-5}\cmidrule(lr){6-8}\cmidrule(lr){9-11}
     & $\ell_2(\times 10^{3}) \downarrow$ & $\mathcal{L}_{\mathrm{mag}} \downarrow$ & $\mathcal{L}_{\mathrm{phs}} \downarrow$ &
    $\mathcal{L}_{\mathrm{MRS}} \downarrow$ & $\boldsymbol{\Delta_{\mathrm{abs}}\theta\downarrow}$ & $\boldsymbol{\Delta_{\mathrm{abs}}\phi\downarrow}$ &
    $\boldsymbol{\Delta_{\mathrm{ang}}\downarrow}$ &PESQ $\uparrow$ &MOS-Q $\uparrow$ & MOS-P $\uparrow$ \\
    \midrule
    \multicolumn{11}{c}{\raisebox{0.6ex}{\textbf{Ablation on Localizer Components}}}\\[0.6ex] 
NoVideo-Renderer  & 1.59 & 0.55 & 1.96 & 2.33   & 0.69 & 0.30 & 0.73 & 2.50  & 3.24  &  2.28 \\
VidEnc-Renderer   & 1.95 & 0.50 & 1.66  &$\textbf{2.11}^{\dagger}$  & $\textbf{0.59}^{\dagger}$ & $\textbf{0.23}^{\dagger}$ & $\textbf{0.66}^{\dagger}$ & 1.78  & 3.03 & $\textbf{3.02}^{\dagger}$ \\
FrozenLoc-Renderer & 2.19 & 0.53 & 1.90  & 2.58    & 0.97 & 0.44 & 1.04 & 1.41 & 2.49 & 2.34 \\
    \midrule
    \multicolumn{11}{c}{\raisebox{0.6ex}{\textbf{Comparison to Renderer Methods}}}\\[0.6ex]
    Localizer-AmbiEnc  & $\textbf{1.55}^{\dagger}$ & $\textbf{0.34}^{\ast}$ & 1.67  & $\textbf{1.69}^{\ast}$    & 0.61 & 0.50 & 0.83 & $\textbf{3.35}^{\dagger}$  & $\textbf{3.68}^{\ast}$& 2.97  \\
    Localizer-Pyroom   & 2.22 & 0.58 & 1.82  & 2.56    & $\textbf{0.58}^{\ast}$ & 0.50 & 0.83 & 1.69 & 2.28  & 2.51  \\
    \midrule
    Ours (NoPT)   & 1.71 & 0.52  &  $\textbf{1.59}^{\ast}$ & 2.46   & 0.84 & 0.36 & 0.92 & 1.48 &2.94  & 2.90 \\
    \textbf{Ours} & $\textbf{1.15}^{\ast}$ & $\textbf{0.44}^{\dagger}$  &  $\textbf{1.59}^{\ast}$ & 2.41   & 0.60 & $\textbf{0.18}^{\ast}$ & $\textbf{0.64}^{\ast}$ & $\textbf{3.42}^{\ast}$ & $\textbf{3.63}^{\dagger}$ & $\textbf{3.16}^{\ast}$\\ 
    \bottomrule
  \end{tabular*}
  \vspace{-10pt} 
\end{table*}

\begin{table}[t]
  \centering
  \caption{\textit{Compatibility comparison with SAG~\cite{360Video2018} on YT-ALL, YT-MUSIC, YT-CLEAN and YT-SPEECH. Lower values indicate better performance (\(\downarrow\)).}}
  \label{tab:baseline}
  \footnotesize
  \setlength{\tabcolsep}{1.5pt}
  \renewcommand{\arraystretch}{1.25}
  \begin{tabular*}{\columnwidth}{@{\extracolsep{\fill}}lcccccccc}
    \hline
      & \multicolumn{2}{c}{\textbf{YT-ALL}} & \multicolumn{2}{c}{\textbf{YT-MUSIC}} & \multicolumn{2}{c}{\textbf{YT-CLEAN}} & \multicolumn{2}{c}{\textbf{YT-SPEECH}}\\
    \cline{2-3}\cline{4-5}\cline{6-7}\cline{8-9}
      & STFT & ENV
      & STFT & ENV
      & STFT & ENV
      & STFT & ENV\\
    \hline
    SAG-NoVideo  & 2.96 & 3.48 & 4.55 & 4.32 & 1.57 & 2.12 & 1.43 & 2.42\\
    SAG            & \textbf{2.78} & \textbf{2.92} & 4.46 & 3.77 & 1.58 & 1.80 & 1.14 & 2.18\\
    Ours-NoVideo & 3.12 & 3.52 & 3.98 & 4.01 & 1.07 & \textbf{1.78} & 1.49 & 1.89\\
    Ours           & 2.91 & 2.99 & \textbf{3.38} & \textbf{3.17} & \textbf{0.91} & \textbf{1.78} & \textbf{0.85} & \textbf{1.71}\\
    \hline
  \end{tabular*}
  \renewcommand{\arraystretch}{1.0}
\end{table}

\subsection{Experimental Setup}
We use YT-SPEECH as the main in-domain benchmark in Table~\ref{tab:foa} and report a protocol-compatibility comparison with SAG on its released datasets for reference in Table~\ref{tab:baseline}.

\textbf{Training Strategy}. We first pretrain the Renderer on a large-scale $360^\circ$ audio-visual dataset from Sphere360~\cite{Omniaudio2025} filtered by an energy-based criterion~\cite{energy_critera}, using AdamW with a learning rate of $5\times10^{-5}$. 
During pretraining,  the Localizer is initialized from a pretrained AVS model and frozen to provide spatial priors, while the Renderer learns the mapping from priors to FOA components across diverse acoustic scenes. 
We then fine-tune the full pipeline on YT-SPEECH with data augmentation for person-aware speech spatialization by jointly optimizing the Prior Head and the Renderer with learning rates $5\times10^{-6}$ and $5\times10^{-5}$, respectively. The loss weights in Eq.~(\ref{eq:loss}) are set to $\lambda_1=1$, $\lambda_2=0.1$, $\lambda_3=0.2$ and $\alpha=0.2$.

\textbf{Data Augmentation}. We apply data augmentation to YT-SPEECH to improve rotational robustness while preserving audio-visual geometric consistency. For each training sample, horizontal yaw rotation and horizontal flipping are applied independently with probabilities 0.8 and 0.2, respectively; the yaw angle is sampled uniformly from $[-\pi,\pi]$. Corresponding FOA channel rotations are applied to maintain spatial alignment.

\subsection{Results}

We conduct comprehensive evaluations using both objective and subjective metrics to measure reconstruction fidelity, spatial accuracy, and speech quality. Reconstruction metrics include $\ell_2$, $\mathcal{L}_{\text{mag}}$, and $\mathcal{L}_{\text{MRS}}$, consistent with the training objective defined in Eq.~(\ref{eq:loss}). 
In addition, we report $\mathcal{L}_{\text{phs}}$ to assess the phase consistency of the generated signals~\cite{NFS}. Spatial metrics employ DOA estimation~\cite{ImmerseDiffusion2024}, reporting the mean absolute azimuth error $\Delta_{\text{abs}}\theta$, elevation error $\Delta_{\text{abs}}\phi$, and overall angular error $\Delta_{\text{ang}}$. Speech quality is evaluated by zeroing the omnidirectional channel for both prediction and reference (i.e., $[0,Y,Z,X]$), decoding FOA to binaural audio using SADIE II KU100 HRIRs, and computing ear-wise PESQ~\cite{PESQ}; we set $W=0$ to avoid PESQ scores being dominated by the ground-truth omnidirectional channel. Subjective evaluation uses MOS-Q and MOS-P to assess overall audio quality and perceived spatial accuracy. We conduct a small-scale listening test with 9 participants; each participant rates 5 clips per method on a 1--5 scale. Scores are reported as the mean across listeners and clips.

Table~\ref{tab:foa} presents our main benchmark results on YT-SPEECH. We compare the full Localizer--Renderer model with several variants.
We ablate the conditioning to the Renderer: NoVideo-Renderer removes the Localizer and uses only audio input, VidEnc-Renderer conditions on video encoder features, and FrozenLoc-Renderer keeps the Localizer fully frozen during training.
We also compare two DOA-based analytic rendering baselines: Localizer-AmbiEnc performs first-order Ambisonics encoding~\cite{Ambisonics}, and Localizer-Pyroom synthesizes FOA signals using Pyroomacoustics~\cite{pyroomacoustics}, both based on the peak direction estimated from the spatial prior map.
Ours (NoPT) disables large-scale pretraining.
Overall, the full model performs best, achieving the lowest $\ell_2$ and $\Delta_{\mathrm{ang}}$ values and the highest PESQ and MOS-P. 
Freezing the Localizer degrades spatial metrics, especially $\Delta_{\mathrm{abs}}\phi$ and $\Delta_{\mathrm{ang}}$, highlighting the need for adaptation.
Compared with the learned Renderer, analytic DOA-based rendering like Localizer-AmbiEnc can obtain strong $\mathcal{L}_{\mathrm{mag}}$ and MOS-Q scores, but remains weaker on spatial errors since it relies on a single peak direction without learned spectral compensation.
Removing pretraining further hurts reconstruction and perceptual quality.

Table~\ref{tab:baseline} provides a compatibility comparison with SAG~\cite{360Video2018} on its three proposed datasets: YT-ALL, YT-MUSIC, and YT-CLEAN together with our curated YT-SPEECH dataset.
Metrics include STFT, the complex short-time Fourier transform distance, and ENV, the Euclidean distance between signal envelopes. 
Following the official SAG protocol, metrics are computed on 1.1~s windows and averaged per 5~s clip.
Overall, our model achieves competitive or improved performance across datasets. 
The most consistent gains are observed on YT-CLEAN and YT-SPEECH, where sources are visually localizable, indicating that our spatial-aware prior and learned renderer generalize well when visual grounding is reliable. 
On YT-MUSIC with multiple mixed sources, our method outperforms SAG in both metrics. 
On YT-ALL, which contains more acoustically diverse content beyond speech, the performance gap is smaller, as the advantage of a speech-aware prior becomes less pronounced.
\begin{figure}[t]
  \centering
  \includegraphics[width=\linewidth]{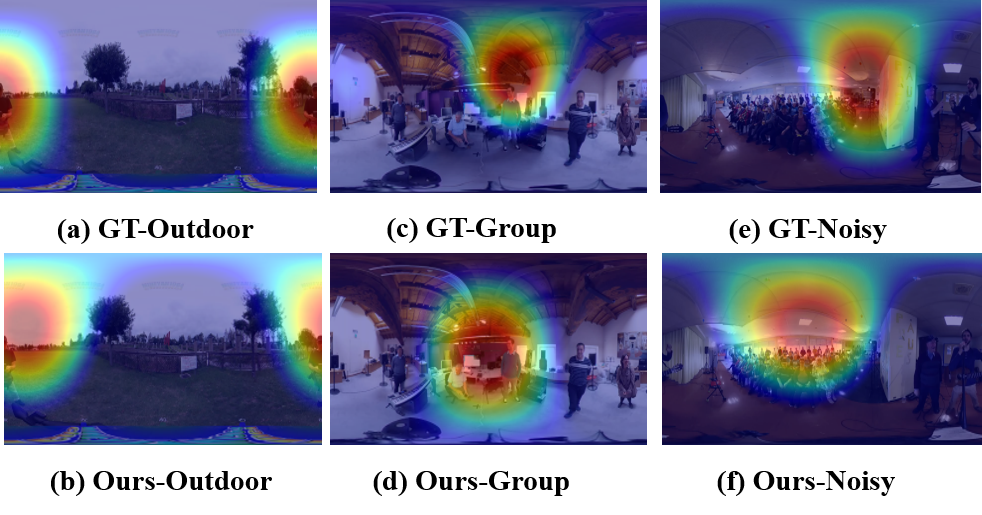}
   \caption{\textit{Qualitative results.} Rows I--III show outdoor, group, and noisy scenes. Ground truth and our prediction are overlaid on the $360^\circ$ frames in (a, c, e) and (b, d, f), respectively.}
  \label{fig:heatmap}
\end{figure}

We also compare the audio energy maps in Fig.~\ref{fig:heatmap} under three challenging conditions selected from our YT-SPEECH dataset. Our model produces sharp and well-aligned activations in outdoor and group-interaction scenes, indicating robust audio--visual grounding. In contrast, noisy backgrounds introduce spurious responses and reduce localization consistency, suggesting that noise remains a primary failure mode.

\section{Conclusion}

We presented YT-SPEECH, a new dataset for the spatialization of speech-dominant $360^\circ$ video scenes with FOA audio, curated from publicly available user-generated content and screened for quality and audio--visual consistency. We further propose a Localizer--Renderer framework, where the Localizer leverages an AVS backbone to generate and interpret frame-wise spatial heatmaps for coherent FOA audio rendering. Experimental results show improved spatial accuracy and speech-related quality metrics. Current limitations include the limited scale of our curated 8.9-hour dataset and reduced stability under overlapping sources and acoustically complex scenes. Future work will explore synthetic or hybrid data with controllable source overlap and ground-truth labels, together with improved augmentation strategies and broader perceptual evaluation.

\section{Acknowledgments}
We thank Pablo Martínez-Nuevo, Sven Ewan Shepstone, and Jon Francombe (Bang \& Olufsen A/S) for valuable discussions. This research was supported by Bang \& Olufsen A/S as part of the AURIC project.

\section{Use of Generative AI Disclosure}
Generative AI tools were used only for language editing and improving the readability of the manuscript. No generative AI tools were used to produce the scientific content, methodology, experiments, results, or conclusions. The authors are fully responsible for the content, claims, methodology, experiments, and conclusions of this paper.

\bibliographystyle{IEEEtran}
\bibliography{mybib}

\end{document}